\documentstyle[preprint,aps,multicol]{revtex}
\newcommand{\be}{\begin{eqnarray}}
\newcommand{\ee}{\end{eqnarray}}
\newcommand{\OP}{\psi}

\begin{document}

\title{Response Functions in Phase Ordering Kinetics}
\author{Gene F. Mazenko }
\address{The James Franck Institute and the Department of Physics\\
The University of Chicago\\
Chicago, Illinois 60637}
\date{\today}
\maketitle
%
%
%
\begin{abstract}
We discuss the behavior of response functions in phase
ordering kinetics within the perturbation theory approach 
developed earlier.  At zeroth order the results
agree with previous gaussian theory calculations.
At second order the 
nonequilibrium exponents $\lambda$
and $\lambda_{R}$  are changed but remain equal.

\end{abstract}


\section{Introduction}

The scaling properties of
response functions for  phase ordering systems
has been the subject of some recent\cite{review} interest.  We
study these properties here using 
the $\phi$-expansion method
developed previously\cite{maz98} to extend perturbation
theory beyond the gaussian level.
We find, while there are no corrections generated 
at second order for the exponent $a$ governing the 
response function,
there are second order corrections for the nonequilibrium
exponents $\lambda$ and $\lambda_{R}$ and
$\lambda =\lambda_{R}$.
We also find the associated scaling function
at second order in perturbation theory.

If $\psi ({\bf r},t)$ is the scalar ordering field 
whose dynamics
are driven by
a time dependent Ginzburg Landau (TDGL) model
(described in detail below), then we can study the
correlation function
\be
C({\bf r}_{1}-{\bf r}_{2},t_{1},t_{2})
=\langle \psi ({\bf r}_{1},t_{1})
\psi ({\bf r}_{2},t_{2})\rangle
~~~.
\ee
This quantity has been studied in a growth kinetics
context\cite{Bray94} for a variety of systems and 
shows the scaling behavior
$C({\bf r},t_{1},t_{2})=
F(r/L(t_{1}),t_{1}/t_{2})$
where $L(t_{1})$ is the characteristic growth law
$L(t_{1})\approx t^{1/z}$
and the growth exponent is $z=2$ for the nonconserved
order parameters studied in this paper.
Focus has been on the onsite correlation function
$C(0,t_{1},t_{2})=
F(0,t_{1}/t_{2})$.
For $t_{1}\gg t_{2}$ we have
$F(0,x)\approx x^{-\lambda/z}$
where $\lambda$ is the well studied\cite{FH88,LM91}  nonequilbrium index.

We can also introduce an external field, $B(1)$, {\it conjugate}
to the order parameter and define the response function
\be
\chi (12)=\left(\frac{\delta \langle \psi (1)\rangle}
{\delta B(2)}\right)_{B=0}
\ee
evaluated at zero field.
For the local (${\bf r}_{1}={\bf r}_{2}$), response function 
it has become customary to write, $t_{1} > t_{2}$,
\be
\chi (0,t_{1},t_{2})=t_{2}^{-(1+a)}f(t_{1}/t_{2})
\label{eq:1.5}
\ee
where for large $x=t_{1}/t_{2}$,
\be
f(x)\approx x^{-\lambda_{R}/z}
~~~.
\label{eq:7}
\ee
The goal in the analysis is to
find the exponents $a$ and $\lambda_{R}$
and the scaling
function $f(x)$.  Here we focus on the regime where both
$t_{1}$ and $t_{2}$ are large enough for the system to be
in the scaling regime.

Let us review some of the results found previously.
The exact result~\cite{Lippiello2000}~\cite{Godreche2000}
for the one-dimensional Ising model gives
$a=0$,
$f(x) \sim (x-1)^{-1/2}$
and $\lambda_{R} =1$.
The solution of the problem in the large $n$ limit
~\cite{Corberi2002} gives the exponent
\be
a=(d-2)/2
~~~,
\label{2.10}
\ee
the scaling function
\be
f(x) \sim \frac{ x^{d/4} }{ (x-1)^{d/2}}
\label{eq:2.12}
~~~,
\ee
and $\lambda_{R}=d/2$.
Berthier, Barrat and Kurchan~\cite{Berthier} have carried
out a gaussian auxiliary function approximation calculation
for this problem.  This has been extended to treat 
one dimension in ~\cite{Corberi2001} with the results:
\be
a=(d-1)/2
\label{eq:2.13}
\ee
and $f(x)$ is again given by Eq.(\ref{eq:2.12}).
Henkel, Pleimling, Godreche and Luck in~\cite{Henkel2001}, 
using conformal
invariance methods, have derived a form
of the scaling function:
\be
f(x) \sim \frac {x^{a+1-\lambda_{R}/z}}{(x-1)^{(a+1)}}
\label{2.14}
~~~.
\ee
This is the same form as Eq.(\ref{eq:2.12}) if
$\lambda_{R}/z=d/4$.

Numerical work\cite{CLZ} has focussed on smoothed integrals of the
fundamental response function $\chi (t_{1},t_{2})$.  
This smoothing procedure
helps with numerical sampling but
can lead to qualitative differences between the 
fundamental response function and its smoothed counter parts
for lower dimensions.

We focus here on the computation
of the local
response $\chi (t_{1},t_{2})=\chi (0,t_{1},t_{2})$
to second order in the $\phi$-expansion developed in 
ref.\onlinecite{maz98} for a scalar order parameter.
At zeroth order we find
\be
\chi_{0}(t_{1},t_{2})=t_{2}^{-1-a}f_{0}(t_{1}/t_{2})
~~~,
\ee
with the exponent a given by Eq.(\ref{eq:2.13})
and
\be
f_{0}(x)=\sqrt{\frac{2}{\pi A_{0} }}
\frac{x^{d/4}}{[4\pi (x-1)]^{d/2}}
\label{eq:18}
\ee
where the constant $A_{0}$ is defined in 
section 4 below.

At second order in perturbation theory we
again find the scaling form given by
Eq.(\ref{eq:1.5}) with $a=\frac{1}{2}(d-1)$
but the scaling function is given now by
\be
f(x)=\sqrt{\frac{2}{\pi A_{0} }}
\frac{x^{\omega -1/2}}{(4\pi )^{d/2}(x-1)^{d/2-v}}
\left(1+\Delta_{d}(x)+\omega^{2}2^{d-1}\bar{g}(x)\right)
\label{eq:17}
\ee
where $\omega$ is determined by Eq.(\ref{eq:170}),
$v$ is given by Eq.(\ref{eq:167}),
$\Delta_{d}(x)$ is given by Eqs.(\ref{eq:140})
and (\ref{eq:141})
and $\bar{g}(x)$ is given by Eq.(\ref{eq:158}) 
and (\ref{eq:164}).
In the large $x$ limit, where $\Delta_{d}(x)$
and $\bar{g}(x)$ go to a constant, we can 
use Eq.(\ref{eq:17}) to identify
the nonequilibrium exponent
\be
\lambda_{R}=d+1-2\omega -2v
~~~.
\label{eq:12}
\ee

Let us use the same self consistent procedures
for evaluating $\lambda_{R}$ as used in I
for $\lambda$ and $\omega$.  We found in I,
at this same order, that
\be
\lambda =\frac{d}{2}+\omega^{2}\frac{2^{d}M_{d}}{3^{d/2+1}}
\label{eq:169}
\ee
with $\omega$ given as the solution to
\be
2\omega+\omega^{2}2^{d}\left(K_{d}+
\frac{M_{d}}{3^{d/2+1}}\right)=1+\frac{d}{2}
\label{eq:170}
~~~,
\ee
\be
M_{d}=\int_{0}^{1}dz \frac{z^{d/2-1}}{[1+z]^{d}}=
\frac{1}{2}\frac{\Gamma^{2}(d/2)}{\Gamma (d)} ~~~.
\ee
and $K_{d}$ is given by Eq.(\ref{eq:165}).
Solving for $\omega$ from Eq.(\ref{eq:170}) and inserting
this into Eq.(\ref{eq:169}) gives the values of $\lambda$
shown in table 1.  Notice, however, that if
we eliminate $K_{d}$ from Eq.(\ref{eq:170}) using
Eq.(\ref{eq:167}) for $v$ and
eliminate  $M_{d}$ in Eq.(\ref{eq:170}) in favor of
$\lambda -d/2$ using Eq.(\ref{eq:169}),
we obtain
\be
2\omega +2v+(\lambda -d/2)=1+d/2
~~~.
\ee
This can be rewritten as
\be
\lambda =1+d-2\omega -2v
~~~.
\ee
Comparing with Eq.(\ref{eq:12})
and we find
\be
\lambda =\lambda_{R}
\ee
and, as was found at lowest order, the nonequilibrium
exponents are equal.  These cancellations, coming from
two very independent calculations of $\lambda$
and $\lambda_{R}$, serve as a severe check on the validity
of the algebra carried out in each.

In the next section we describe the setting up of the
perturbation theory used to obtain these results.

\begin{table}
\begin{tabular}{|c|c|c|}
\hline
& &\\
$dimension$ &  $\lambda =\lambda_{R} $   &  $\omega $ \\
& &\\
$1$      & $0.6268..$ &  $0.4601..$     \\
$2$      & $1.1051..$ & $0.6877..$  \\
$3$      & $1.5824..$ &  $0.9067..$  \\
$large $ & $ d/2 $    & $\frac{d}{2}(\sqrt{2}-1)$ \\
 &  & \\
\hline
\end{tabular}
\caption{Second order values for the exponents 
$\lambda =\lambda_{R} $ and the parameter $\omega$ from 
ref.(\protect\onlinecite{maz98}).}
\end{table}

\section{Introduction of Ordering Auxiliary Field}

Consider a system
where the ordering kinetics
are driven by the simplest nonlinear 
time dependent Ginzburg-Landau (TDGL) model.
Assuming that we quench a system from a disordered 
high temperature
state to zero temperature, where the thermal noise 
is set to zero,
we have the equation of motion.
\be
\frac{\partial \psi}{\partial t}=\Gamma\left(- V'(\psi)+c\nabla^{2}\psi
+B\right) +\delta (t_{1}-t_{0})\psi _{0}~~~.
\label{eq:22}
\ee
where $V(\psi )$ is the driving potential.  Typically this is chosen
for simplicity to be of the $\psi^{4}$ type:
$V=\frac{1}{4}(1-\psi^{2})^{2}$.
We have also included an external
field $B({\bf r},t)$, conjugate to the order parameter,
 in the equation of motion.  Choosing units of
time and space such that $\Gamma =c=1$ we can write
\be
\Lambda (1) \psi (1)=-V'[\OP (1)] +B(1)+\delta (t_{1}-t_{0})\psi _{0}
\label{eq:5}
\ee
where the diffusion operator
\be
\Lambda (1)=\frac{\partial}{\partial t_{1}}-\nabla^{2}_{1}
\ee
is introduced along with the short-hand notation that $1$ denotes
$({\bf r_{1}},t_{1})$.

We show, extending the analysis in I to include an external
field, that the order parameter $\psi$ can be divided into an
ordering component $\sigma$ and an equilibrating component
$u$,
\be
\psi =\sigma [m] + u[m]
\label{eq:30b}
\ee
where $\sigma [m]$ is  the solution to the Euler-Lagrange
equation
\be
\frac{d^{2}\sigma }{dm^{2}}\equiv \sigma_{2}=V'[\sigma [m]] 
\label{eq:EL}
\ee
connecting the degenerate states 
$\psi =\pm \psi_{0}$, $V^{\prime}(\pm \psi_{0})=0$.
In this equation $m$ is taken to be the coordinate.
In the case of a $\psi^{4}$  potential we have the solution
\be
\sigma [m]=tanh(\frac{1}{\sqrt{2}}m)
~~~.
\ee
It is shown in ref.\onlinecite{maz98} that 
the equilibrating field $u$ 
decays exponentially to zero at long times.  The scaling properties
of the theory are carried by the ordering field
$\sigma [m]$.

It was shown in I that the theory is self-consistent if
the auxiliary field saisfies an equation of motion of the 
the form
\be
\Lambda (1)m(1)=\Xi (1)+B(1)  
+\delta (t_{1}-t_{0})m _{0}({\bf r}_{1})~~~.
\label{eq:27}
\ee
where $\Xi$ is a function of $m$ which must
scale as $\approx 1/L(t)$ in the scaling regime
and self-consistently generates ordering.  In particular
$\Xi$ must be such that 
$\langle m^{2}(1)\rangle \approx L^{2}(t_{1})$.
In I and here we study the nonlinear model where
\be
\Xi (1) =\xi (t_{1})\sigma (m(1))
\label{eq:28}
\ee
and, if scaling is to hold, we find self-consistently
that
$\xi (t_{1})\approx 1/L(t_{1})$ and is independent of
the field $m$.  

\section{Field Theory for Auxiliary Field}

Let us consider the field theory associated with 
the equation of motion for $m({\bf r},t)$ given by
Eqs.(\ref{eq:27}) and (\ref{eq:28}).
Our  development will follow  the standard
Martin-Siggia-Rose\cite{MSR} method in its functional integral form
as developed by
DeDominicis and Peliti\cite{DP}.  
In the MSR method the field
theoretical development requires a doubling of operators to include the
field $M$ which is conjugate to $m$.
We give here an overview of the development in I needed
here to treat the response function including the
coupling to an external field.

Following
standard procedures, described in more detail in I, 
averages of interest are given as functional
integrals over the
fields $m$, $M$ and weighted by the probability
distribution $P[m,M]$:
\be
\langle f(m,M)\rangle =
\int {\cal D}m {\cal D}M P[m,M] f(m,M)
\label{eq:27a}
\ee
\be
P[m,M]=e^{A_{T}(m,M)}/Z(H,h)
\nonumber
\ee
and
\be
Z(h,H)=\int {\cal D}m {\cal D}Me^{A_{T}(m,M)} ~~~.
\ee
The action takes the form
\be
A_{T}(m,M)=A(m,M)+\int d1 \left[h(1)m(1)+H(1)M(1)\right]
\ee
where
\be
A(m,M)=-i\int d1M(1)
\left[\Lambda (1)
m(1)-\xi(1)\sigma (1) -B(1)\right]
\nonumber
\ee
\be
-\frac{1}{2}\int ~d1\int ~d2~M(1)\Pi_{0}(12)M(2)
\label{eq:30}
\ee
with 
\be
\Pi_{0} (12) \equiv \delta (t_{1}-t_{0})\delta (t_{1}-t_{2})
g({\bf r}_{1}-{\bf r}_{2}) ~~~.
\label{eq:33}
\ee
In these equations we use the notation,
$\int ~d1=\int ~dt_{1}d^{d}r_{1}$~,
and assume that the initial field $m_{0}({\bf r})$
is gaussian and has a variance given by
\be
<m_{0}({\bf r}_{1})m_{0}({\bf r}_{2})>=g({\bf r}_{1}-{\bf r}_{2}) ~~~.
\ee
We can generate 
correlation functions as functional derivatives in terms of sources 
$h$ and $H$ which
couple to the conjugate fields.  

The fundamental equations of motion are given by the identities
\be
<\frac{\delta}{\delta M(1)}A_{T}(m,M)>_{h}=0
\label{eq:33a}
\ee
\be
<\frac{\delta}{\delta m(1)}A_{T}(m,M)>_{h}=0
\label{eq:33b}
\ee
where the subscript $h$ indicates that the average includes the
source fields $h$ and $H$.  Taking the derivative
in Eq.(\ref{eq:33a}) we obtain
\be
i\left[\Lambda (1)
<m(1)>_{h}-Q_{1}(1)\right]
\nonumber
\ee
\be
=-\int d2 ~\Pi_{0} (12)<M(2)>_{h}+H(1)-iB(1)
\label{eq:37}
\ee
where the nonlinearities are included in
\be
Q_{1}(1)=\xi(1)<\sigma (1)>_{h}
\label{eq:38a}
\ee
Eq.(\ref{eq:33b}) gives
\be
-i\left[\tilde{\Lambda}(1)\langle M(1)\rangle
+\hat{Q}_{1}(1)
\right]=h(1)
\nonumber
\ee
where 
\be
\tilde{\Lambda}(1)=\frac{\partial}{\partial t_{1}}+\nabla^{2}_{1}
\nonumber
\ee
and 
\be
\hat{Q}_{1}(1)=\xi(1)\langle \sigma_{1} (m(1))M(1)\rangle
~~~.
\ee

Clearly we can go on and generate equations for all of the cumulants by
taking functional derivatives
of Eqs.(\ref{eq:37}) and (\ref{eq:38a}).
Let us introduce the notation that
$G_{A_{1},A_{2},....,A_{n}}(12...n)$ is the $n^{th}$ order cumulant
for the set of fields $\{A_{1},A_{2},....,A_{n}\}$ where field
$A_{1}$ has argument $(1)$, field $A_{2}$ has argument $(2)$, et cetera.
This notation is needed when we mix cumulants with $m$ and $M$.
As an example
\be
G_{Mmmm}(1234)=\frac{\delta^{3}<m(4)>_{h}}{\delta H(1)
\delta h(2)\delta h(3)} ~~~.
\ee
As a short hand for cumulants involving only $m$ fields we write
\be
G_{n}(12\cdots n)=\frac{\delta^{n-1}}{\delta h(n)\delta h(n-1)
\cdots \delta h(2)}<m(1)>_{h} ~~~.
\ee
The equations governing the $n^{th}$ order cumulants are given by
\be
-i\left[\tilde{\Lambda}(1)G_{Mm...m}(12...n)
+\hat{Q}_{n}(12...n)\right]=0
\label{eq:82}
\ee
and
\be
i\left[\Lambda (1)G_{n}(12...n)-Q_{n}(12...n)\right]
=-\int
d\bar{1} ~\Pi_{0} (1\bar{1})G_{Mm...m}(\bar{1}2...n) ~~~.
\label{eq:83}
\ee
The $Q's$ are defined by
\be
\hat{Q}_{n}(12...n)=
\frac{\delta^{n-1}}{\delta h(n)\delta h(n-1)\cdots\delta h(2)}
\hat{Q}_{1}(1)
\label{eq:88}
\ee
\be
Q_{n}(12...n)=
\frac{\delta^{n-1}}{\delta h(n)\delta h(n-1)\cdots\delta h(2)}
Q_{1}(1).
\label{eq:89}
\ee
With this notation the equations determining the 
two-point functions can be
written as
\be
-i\left[\tilde{\Lambda}(1)G_{Mm}(12)
+\hat{Q}_{2}(12)\right]=\delta (12)
\label{eq:44}
\ee
\be
i\left[\Lambda (1)G_{2}(12)-Q_{2}(12)\right]
=-\int
d\bar{1} ~\Pi_{0} (1\bar{1})G_{Mm}(\bar{1}2)
~~~.
\label{eq:45}
\ee

This is all formally exact.  In order to develop
perturbation theory the next step is
to show that $Q_{1}(1)$ and
$\hat{Q}_{1}(1)$  can be expressed in terms of the
singlet probability distribution
\be
P_{h}(x,1)=<\delta (x-m(1))>_{h}
~~~.
\ee
One finds
\be
Q_{1}(1)=\int dx ~
\xi (1)\sigma (x)~P_{h}(x,1) 
\ee
\be
\hat{Q}_{1}(1)=\int ~dx ~ \xi (1)\sigma_{1} (x)
\left[<M(1)>_{h}
+\frac{\delta}{\delta H(1)}\right]
P_{h}(x,1).
\label{eq:49a}
\ee
Then any perturbation theory expansion for $P_{h}(x,1)$ will
lead immediately to an expansion for $\hat{Q}_{1}(1)$ and
$Q_{1}(1)$.  We can then obtain  $\hat{Q}_{n}$ and $Q_{n}$ by functional
differentiation.

The development of a perturbation theory expansion 
for $P_{h}(x,1)$ begins by using the
integral representation for the $\delta$-function:
\be
P_{h}(x,1)= \int\frac{dk}{2\pi}e^{-ikx}\Phi (k,h,1)
\label{eq:50}
\ee
where
\be
\Phi (k,h,1)=
\langle e^{ikm(1)}\rangle_{h}
~~~.
\ee
The average of the exponential is precisely of the form which can be
rewritten in terms of cumulants:
\be
\Phi (k,h,1)
=exp\left[\sum_{s=1}^{\infty}\frac{(ik)^{s}}{s!}G_{s}(11...1)\right] 
\label{eq:103}
\ee
where $G_{s}(11...1)$ is the s-order cumulant for the
field $m(1)$.

Consider
first the lowest-order contribution to $Q_{n}$ which does not vanish with the
external fields $h,H$:
\be
Q_{2}(12)=\int dx ~
\xi (1)\sigma (x)\frac{\delta}{\delta h(2)}P_{h}(x,1) ~~~.
\ee
We have shown that the
$n^{th}$ order cumulants are
of order $\frac{n}{2}-1$ in an expansion parameter we will develop.
Expanding $\Phi (k,h,1)$ in powers of the cumulants with $n > 2$ and
keeping
terms up to the 4-point cumulant, we obtain
\be
P_{h}(x,1)=\left[1-\frac{1}{3!}G_{3}(111)\frac{d^{3}}{dx^{3}}
+\frac{1}{4!}G_{4}(1111)\frac{d^{4}}{dx^{4}}+\cdots\right]
P_{h}^{(0)}(x,1)
\ee
where
\be
P_{h}^{(0)}(x,1)=\int\frac{dk}{2\pi} \Phi_{0}(k,h,1)e^{-ikx}
\ee
and
\be
\Phi_{0}(k,h,1)=e^{ikG_{1}(1)}
e^{-\frac{1}{2}k^{2}G_{2}(11)} ~~~.
\ee
Then,
after taking the derivative with
respect to $h(2)$, setting the external fields to zero, and
neglecting all cumulants with $n > 2$, we obtain
\be
\Phi_{0}(k,h=0,1)=e^{-\frac{1}{2}k^{2}S_{2}(1)} ~~~,
\ee
and
\be
Q_{2}^{(0)}(12)=\int dx ~
\xi (1)\sigma (x)
\int\frac{dk}{2\pi}e^{-ikx}ik G_{2}(12)e^{-\frac{1}{2}k^{2}S_{2}(1)}
\nonumber
\ee
where we have defined in zero external field
\be
S_{2}(1)\equiv G_{2}(11)=<m^{2}(1)> ~~~.
\ee
In the scaling regime, where $S_{2}(1)$ is very large,
we can replace $\sigma (x)\rightarrow \psi_{0}~ sgn (x)$
in the integral and obtain:

\be
Q_{2}^{(0)}(12)=\xi (1)\psi_{0}G_{2}(12)
\int dx
~ sgn (x)
\int\frac{dk}{2\pi}ik e^{-ikx} e^{-\frac{1}{2}k^{2}S_{2}(1)}
\nonumber
\ee
\be
=\xi (1)\psi_{0}G_{2}(12)\int dx
{}~sgn (x)\left[-\frac{d}{dx}\Phi_{0}(x,1)\right] .
\nonumber
\ee
In this case we integrate by parts in the integral over $x$ and use
$\frac{d}{dx} sgn (x) =2\delta (x)$
to obtain
\be
Q_{2}^{(0)}(12)=\xi (1)\psi_{0}G_{2}(12)
2\Phi_{0}(0,1)
\nonumber
\ee
\be
=\xi (1)\psi_{0}G_{2}(12)\sqrt{\frac{2}{\pi S_{2}(1)}}~~~.
\ee

Turning to $\hat{Q}_{2}(12)$ we note that it is given
by taking the derivative of Eq.(\ref{eq:49a}) with
respect to $h(2)$.  In the scaling regime where
the characteristic length $L(t)$ is large we can
replace $\sigma_{1} (x) \rightarrow \psi_{0}2\delta (x)$.
Then we need only consider 
\be
\hat{Q}_{2}(12)=\xi (1)\psi_{0}\int ~dx  2\delta (x)
\nonumber
\ee
\be
\times\left[G_{Mm}(12)+<M(1)>_{h}
\frac{\delta}{\delta h(2)}
+\frac{\delta^{2}}{\delta h(2)\delta H(1)}\right]
P_{h}(x,1)
\nonumber
~~.
\ee

A key observation is that as we analyze contributions to
$Q_{n}$ or $\hat{Q}_{n}$ we will find that each term consists of
products of correlation functions and response functions with
legs tied together by factors defined by
\be
\phi_{p}(1)\equiv\int dx~sgn (x) \int \frac{dk}{2\pi}
ik^{2p+1}e^{-ikx}\Phi_{0}(k,1)
\ee
\be
=2\int \frac{dk}{2\pi}k^{2p}e^{-\frac{1}{2}k^{2}S_{2}(1)}
=\left(-2\frac{d}{dS_{2}(1)}\right)^{p}\phi_{0}(1)
\nonumber
\ee
where we have used an integration by parts in going from the
first to the second line and defined
\be
\phi_{0}(1)=2\int \frac{dk}{2\pi}e^{-\frac{1}{2}k^{2}S_{2}(1)}
=\sqrt{\frac{2}{\pi S_{2}(1)}}~~~.
\label{eq:61}
\ee
Each term in the perturbation theory expansion for
$Q_{n}$ or $\hat{Q}_{n}$
will be
proportional to factors of $\phi_{p}$.
The perturbation expansion is ordered by the
sum of the labels $p$ on $\phi_{p}$.  Thus a contribution with insertions
$\phi_{1}\phi_{2}\phi_{1}$ ~,
each factor
typically associated with different times,
is of ${\cal O}(4)$.
We refer to this expansion as the {\it $\phi$-expansion}.
It should be emphasized that at this stage that this is a {\it formal}
expansion.  At order $n$ it is true that
$\phi_{p}\approx L^{-(2p+1)}$
which is small, however it will be multiplied, depending on the quantity
expanded, by positive factors of $L(t)$ such that each term in the
expansion in $\phi_{p}$ has the same
overall leading power with respect to $L(t)$.

To see how this expansion works let us consider first the two-point
quantity $Q_{2}(12)$, defined by
\be
Q_{2}(12)=\int~dx~\xi (1)\psi_{0} sgn(x)
\frac{\delta P_{h}(x,1)}{\delta h(2)} ~~~.
\ee
Using Eqs.(\ref{eq:50}) and (\ref{eq:103}) and taking the derivatives
with respect to $h(2)$, it was shown in I,
in the case of zero external
fields, that
\be
Q_{2}(12)=\psi_{0}\xi(1)\int~dx ~sgn(x)\int~
\frac{dk}{2\pi}e^{-ikx}\Phi (k,h=0,1)
\nonumber
\ee
\be
\times\sum_{s=0}^{\infty}\frac{(ik)^{2s+1}}{(2s+1)!}
G_{2s+2}(11...12) ~~~.
\ee
Since all odd cumulants vanish in
the case of zero external fields  we have
\be
\Phi (k,h=0,1)=exp\left[\sum_{s=1}^{\infty}\frac{(-1)^{s}k^{2s}}{(2s)!}
S_{2s}(1)\right]
\ee
where
\be
S_{2s}(1)=G_{2s}(11...1) ~~~.
\ee
Let us define the set of vertices
\be
{\cal V}_{p}(1)=
\int dx~sgn (x) \int \frac{dk}{2\pi}
ik^{2p+1}e^{-ikx}\Phi(k,h=0,1) ~~~,
\label{eq:141a}
\ee
which can be written, after following the same set of 
steps in reducing
the original expression for $\phi_{p}$, to
\be
{\cal V}_{p}(1)=2 \int \frac{dk}{2\pi} k^{2p}\Phi(k,h=0,1)
\ee
which is independent of position.  Then the  quantity $Q_{2}(12)$,
which appears in the
equation of motion for $G_{2}(12)$, is given in the form
\be
Q_{2}(12)
=\psi_{0}\xi(1) \sum_{s=0}^{\infty}\frac{(-1)^{s}}{(2s+1)!}
{\cal V}_{s}(1)
G_{2s+2}(11...12)
\label{eq:131}
\ee
where we have used the definition of ${\cal V}_{s}(1)$ given by
Eq.(\ref{eq:141a})
in the last step.

It should be clear that the vertices ${\cal V}_{s}(1)$ are of at least
${\cal O}(s)$ in the $\phi$-expansion.  By direct expansion of
$\Phi (k,h=0,1)$ about $\Phi_{0} (k,h=0,1)$ we obtain
\be
{\cal V}_{s}(1)=\phi_{s}(1)+\frac{S_{4}(1)}{4!}\phi_{s+2}(1)
-\frac{S_{6}(1)}{6!}\phi_{s+3}(1) +\cdots ~~~.
\label{eq:69}
\ee
It was found self-consistently in I that $\ell^{th}$ order
cumulants, like $S_{\ell}(1)$, are of
${\cal O}(\frac{\ell}{2}-1)$.
The terms in the expansion for ${\cal V}_{s}$,
given by Eq.(\ref{eq:69}), are of ${\cal O}(s)$,
${\cal O}(s+3)$, and  ${\cal O}(s+5)$ respectively.

Let us turn next to $\hat{Q}_{2}(12)$.
It was shown in I, 
in the same limit of zero applied field,
that the nonlinear contribution to the
equation of motion for $G_{mM}$ can be written as
\be
\hat{Q}_{2}(12)
=\xi (1)\psi_{0}\sum_{s=0}^{\infty}\frac{(-1)^{s}}{(2s)!} {\cal V}_{s}(1)
G^{(2s+2)}_{mm...mMm}(11...112 )  ~~~.
\label{eq:70}
\ee

Before going on to discuss the perturbation theory
calculation of the physical response function
let us made sure the theory is sensible at zeroth order
where, from Eqs.(\ref{eq:131}) and (\ref{eq:70})
\be
Q_{2}^{(0)}(12)=\xi(1)\psi_{0}\phi_{0}(1)G_{2}(12)
\equiv \omega_{0} (1)G_{2}(12)
\label{eq:71a}
\ee
\be
\hat{Q}_{2}^{(0)}(12)=\xi(1)\psi_{0}\phi_{0}(1)G_{Mm}(12) 
\equiv \omega_{0} (1)G_{Mm}(12)
\label{eq:72a}
\ee
where for a scaling solution 
\be
\omega_{0} (1)=\xi(1)\psi_{0}\phi_{0}(1)
\ee
must fall off as $1/t_{1}$ for large $t_{1}$.

\section{Gaussian Theory}

Inserting Eqs.(\ref{eq:71a}) and (\ref{eq:72a})
into Eqs.(\ref{eq:44}) and (\ref{eq:45})
we obtain the equations for the response function:
\be
-i\left[\tilde{\Lambda} (1)+\omega_{0}(1)\right]G^{(0)}_{Mm}(12)
=\delta (12)
\ee
and the correlation function
\be
i\left[\Lambda (1)-\omega_{0}(1)\right]G^{(0)}_{2}(12)=
-\int ~d{\bar 1}\Pi_{0}(1\bar{1})G^{(0)}_{Mm}(\bar{1}2)
~~~.
\ee
It is not difficult to show that
\be
i\left[\Lambda (1)-\omega_{0}(1)\right]G^{(0)}_{mM}(12)
=\delta (12)
~~~.
\label{eq:75}
\ee
Using this last result we have that the correlation function
can be written as
\be
G^{(0)}_{2}(12)
=\int d\bar{1}\int d\bar{2}iG^{(0)}_{mM}(1\bar{1})
iG^{(0)}_{mM}(2\bar{2})\Pi_{0}(\bar{1}\bar{2})
\ee
where $\Pi_{0}(12)$  is given by Eq.(\ref{eq:33}).

The first step in the construction of the 
solution to these equations is to  Fourier transform
Eq.(\ref{eq:75})
over space:
\be
\left[\frac{\partial}{\partial t_{1}}+q^{2}-\omega_{0}(t_{1})\right]
iG^{(0)}_{mM}(q,t_{1}t_{2})
=\delta (t_{1}-t_{2}) ~~~.
\ee
This first-order differential equation has the solution
\be
iG^{(0)}_{mM}(q,t_{1}t_{2})=\theta (t_{1}-t_{2})
exp\left[\int_{t_{2}}^{t_{1}}d\tau
\left(-q^{2}+\omega_{0} (\tau )\right)\right]
\ee
\be
=\theta (t_{1}-t_{2})R(t_{1},t_{2})e^{-q^{2}(t_{1}-t_{2})}
\nonumber
\ee
and we have defined
\be
R(t_{1},t_{2})=e^{\int_{t_{2}}^{t_{1}}~d\tau\omega_{0} (\tau )}  ~~~.
\label{eq:161}
\ee
Taking the inverse Fourier transform
we obtain
\be
iG^{(0)}_{mM}(r,t_{1}t_{2})=\theta (t_{1}-t_{2})R(t_{1},t_{2})
\frac{e^{-\frac{r^{2}}{4(t_{1}-t_{2})}}}{[4\pi (t_{1}-t_{2})]^{d/2}}
~~~.
\label{eq:84}
\ee

Let us turn our attention to the correlation function:
Taking the Fourier transform and
inserting the results for the propagators and $\Pi_{0}$ we obtain
\be
G^{(0)}_{2}(q,t_{1}t_{2})=\theta (t_{1}-t_{0})\theta (t_{2}-t_{0})
R(t_{1},t_{0})R(t_{2},t_{0})e^{-2q^{2}T}\tilde{g}(q)
\label{eq:85}
\ee
where $\tilde{g}(q)$ is the Fourier transform of the initial
correlation function
and
$T=\frac{t_{1}+t_{2}}{2}-t_{0}$.
While we are primarily interested in the long-time scaling
properties of our system, we can retain some control over the
influence of initial conditions and still be able to carry out the
analysis analytically if we introduce the initial condition
\be
\tilde{g}(q)=g_{0}e^{-\frac{1}{2}(q\ell )^{2}}
\ee
or
\be
g( r)=g_{0}
\frac{e^{-\frac{1}{2}(r/\ell )^{2}}}{(2\pi \ell^{2})^{d/2}} ~~~.
\ee
Inserting this form into Eq.(\ref{eq:85}) and doing the wavenumber
integration we obtain
\be
G^{(0)}_{2}({\bf r},t_{1}t_{2})
=R(t_{1},t_{0})R(t_{2},t_{0})
\frac{g_{0}}{[2\pi (\ell^{2}+4T)]^{d/2}}
e^{-\frac{1}{2}r^{2}/(\ell^{2}+4T)} .
\label{eq:83a}
\ee

Let us turn now to the quantity $R(t_{1},t_{2})$ defined by
Eq.(\ref{eq:161}).   We assume that $\omega_{0}$ has the form
given by
\be
\omega _{0}(1)=\frac{\omega}{t_{1}+t_{c}}
~~~,
\label{eq:73}
\ee
where $t_{c}$ is a parameter (or function of
$t$ which goes to a value  $t_{c}$ for large times) such
that the correlations of $m$ has a smooth early
time behavior.
$\omega$ is a constant we will determine.
Evaluating the integral
\be
\int_{t_{2}}^{t_{1}}~d\tau~ \omega_{0}(\tau )=
\int_{t_{2}}^{t_{1}}~d\tau\frac{\omega}{t_{c}+\tau }
=\omega~ln\left(\frac{t_{1}+t_{c}}{t_{2}+t_{c}}\right) ~~~,
\ee
we obtain
\be
R(t_{1},t_{2})=\left(\frac{t_{1}+t_{c}}{t_{2}+t_{c}}\right)^{\omega}~~~.
\ee
Inserting this result back into Eq.(\ref{eq:83a}) leads to the
expression
for the correlation function
\be
G^{(0)}_{2}(r,t_{1}t_{2})=g(0)
\left(\frac{t_{1}+t_{c}}{t_{0}+t_{c}}\right)^{\omega}
\left(\frac{t_{2}+t_{c}}{t_{0}+t_{c}}\right)^{\omega}
\frac{e^{-r^{2}/8T}}{(8\pi T)^{d/2}}~~~.
\ee
If we are to have a self-consistent scaling equation then the
autocorrelation function $(r=0)$, at large equal times
$t_{1}=t_{2}=t$, given by
\be
S_{2}^{(0)}(t)
=t^{2\omega-d/2}\frac{1}{(t_{0}+t_{c})^{2\omega}}
\frac{g_{0}}{(8\pi )^{d/2}} ~~~,
\ee
must have the form $S_{2}^{(0)}(t)=A_{0}t$.  Comparing we see the
exponent $\omega$ must be given by
\be
\omega=\frac{1}{2}(1+\frac{d}{2})
\label{eq:94}
\ee
and the amplitude by
\be
A_{0}=\frac{1}{(t_{0}+t_{c})^{2\omega}}
\frac{g_{0}}{(8\pi )^{d/2}}
~~~.
\label{eq:95}
\ee

The general expression for the correlation function
can be rewritten in the convenient form
\be
G^{(0)}_{2}(r,t_{1}t_{2})=\sqrt{S^{(0)}_{2}(t_{1})S^{(0)}_{2}(t_{2})}
\Phi_{0}(t_{1}t_{2})
e^{-\frac{1}{2}r^{2}/(\ell^{2}+4T)}
\ee
where
\be
\Phi_{0}(t_{1}t_{2})=\left(\frac{\sqrt{(t_{1}+t_{c})
(t_{2}+t_{c})}}
{ T+t_{c}+t_{0}}
\right)^{d/2}~~~.
\ee
The nonequilibrium exponent is defined in the long-time limit by
\be
\frac{G^{(0)}_{2}(0,t_{1}t_{2})}
{\sqrt{S_{0}(t_{1})S_{0}(t_{2})}}=
\left(\frac{\sqrt{(t_{1}+t_{c})(t_{2}+t_{c})}}
{T+t_{0}+t_{c}}\right)^{\lambda}
\label{eq:178}
\ee
and we obtain the OJK\cite{OJK,Bray94}  result
\be
\lambda =\frac{d}{2}~~~.
\ee

Looking at equal times we have that
\be
f_{0}(x)=\frac{G^{(0)}_{2}(r,tt)}{S^{(0)}_{2}(t)}=e^{-x^{2}/2}
\ee
where the scaled length is defined by
${\bf x}={\bf r}/4t$~.
$f_{0}(x)$
is just the well known OJK result for the scaled auxiliary correlation
function.  The connection to the physical order parameter correlation
function is discussed in I.  We will not need these results here.

\section{Nonlinear Response}

\subsection{General Expansion}

We are interested in the nonlinear response function
\be
\chi (12)=\frac{\delta}{\delta B(2)}
 \langle \sigma (1)\rangle_{h,B=0}
\nonumber
\ee
\be
=i\langle \sigma (1)M(2)\rangle_{h,B=0}
\nonumber
\ee
\be
=i\frac{\delta}{\delta H(2)}\langle \sigma (1)\rangle_{h,B=0}
\ee
where the second line follows from Eqs.(\ref{eq:27a})
and (\ref{eq:30}).
We have now set $B(1)=0$ and we can express the reponse function
in terms of a functional derivative of the singlet 
probability distribution function just as in the computation
of $Q_{2}$:
\be
\chi (1,2)
=i\frac{\delta}{\delta H(2)}
\int dx~ \sigma (x)P_{h}(x,1)
~~~.
\ee

The perturbation theory expansion for $P_{h}(x,1)$ was
discussed in section 3.  Using Eqs.(\ref{eq:50})
and (\ref{eq:103}) we have
\be
\chi (1,2)
=i\frac{\delta}{\delta H(2)}
\int dx~ \sigma (x)
\int\frac{dk}{2\pi}e^{-ikx}\Phi (k,h,1)
\nonumber
\ee
\be
=i\int dx ~\sigma (x)
\int\frac{dk}{2\pi}e^{-ikx}\Phi (k,h,1)
\sum_{s=1}^{\infty}\frac{(ik)^{s}}{s!}G_{s,M}(11...12)
~~~.
\ee
We can then set the source fields to zero and obtain
\be
\chi (1,2)
=i\sum_{s=0}^{\infty}\frac{(-1)^{s}}{(2s+1)!}
{\cal V}_{s}(1)
G_{2s+1,M}(11...12)
\ee
where we introduce the same vertices
${\cal V}_{s}$ as in section 3.
We have the exlicit expansion for the physical
response function to the lowest two orders:
\be
\chi (1,2)
=i\phi_{0}(1)G_{mM}(12) +i\phi_{1}(1)\frac{(-1)}{3!}
G_{mmmM}(1112)+\ldots
~~~.
\ee
The leading term says that the physical response
contains a term proportional to the MSR response
function $G_{mM}(12)$.

\subsection{Zeroth Order}

At zeroth order we have the response function
\be
\chi_{0}(1,2)=i\phi_{0}(1)G^{(0)}_{mM}(12)
~~~.
\ee
$\phi_{0}(1)$ is given by Eq.(\ref{eq:61}) and
$iG^{(0)}_{mM}(12)$ is given by Eq.(\ref{eq:84}).
Putting these together we obtain
\be
\chi_{0}(1,2)=
\frac{2}{\sqrt{2\pi S^{(0)}_{2}(1)}}
\theta (t_{1}-t_{2})R(t_{1},t_{2})
\frac{e^{-\frac{r^{2}}{4(t_{1}-t_{2})}}}{[4\pi (t_{1}-t_{2})]^{d/2}}
~~~.
\label{eq:110}
\ee
If we focus on the on- site response function:
\be
\chi_{0} (t_{1},t_{2})=
\chi_{0} (0,t_{1},t_{2})
=\frac{2}{\sqrt{2\pi S^{(0)}_{2}(1)}}
\theta (t_{1}-t_{2})
\left(\frac{t_{1}+t_{c}}{t_{2}+t_{c}}\right)^{\omega}
\frac{1}{[4\pi (t_{1}-t_{2})]^{d/2}}
~~~.
\ee
If we  assume $t_{1} > t_{2}$
and write $S^{(0)}_{2}(1)=A_{0}t$, we have
\be
\chi_{0} (t_{1},t_{2}) = \sqrt{\frac{2}{\pi A_{0} t_{1}}}
\left(\frac{t_{1}+t_{c}}{t_{2}+t_{c}}\right)^{\omega}
\frac{1}{[4\pi (t_{1}-t_{2})]^{d/2}}
~~~.
\ee
It is conventional to write this in the scaling form
\be
\chi_{0} (t_{1},t_{2})=t_{2}^{-1-a}f_{0}(x)
\ee
where $x=t_{1}/t_{2}$.  Clearly we can identify
$1+a = 1/2+d/2$, or
\be
a=\frac{d-1}{2}
\ee
and the scaling function is given by
\be
f_{0}(x)=\sqrt{\frac{2}{\pi A_{0} }}
\frac{x^{\omega -1/2}}{[4\pi (x-1)]^{d/2}}
\ee
and, using Eq.(\ref{eq:94}) for $\omega$,
we find in the large $x$ limit the form
given by Eq.(\ref{eq:7}), with
\be
\lambda_{R}=d+1-2\omega =d/2=\lambda
~~~.
\ee

\subsection{Second Order}

At second-order we have two contributions.
The second-order contibution to
$G_{mM}$ can be read off from Eq.(165)
in I with the result
\be
\chi^{(2)}_{1}(12)=i\phi_{1}(1)G^{(2)}_{mM}(12)
\ee
where
\be
G^{(2)}_{mM}(12)
=\int d\bar{1}d\bar{2} G^{(0)}_{mM}(1\bar{1})
\Sigma^{(2)}(\bar{1}\bar{2})
G^{(0)}_{mM}(\bar{2}2)
\ee
and the associated self-energy is given by
\be
\Sigma^{(2)}(12)=\frac{1}{2}\left(-i\omega_{1}(1)\right)
G^{(0)}_{mM}(12)
\left(G^{(0)}_{2}(12)\right)^{2}
\left(-i\omega_{1}(2)\right)
~~~.
\ee
$\omega_{1}$ is given by
\be
\omega_{1}(1)=\frac{\omega_{0}(1)}{S^{(0)}_{2}(1)}
~~~.
\ee

The second contribution at second order is given by
\be
\chi^{(2)}_{2}(12)=i\phi_{1}(1)\frac{(-1)}{3!}G^{(0)}_{mmmM}(1112)
~~~.
\ee
We therefore need the lowest order expression for
$G^{(0)}_{mmmM}(1112)$.  The lowest order expressions for
 all of the various four-point cumulants were worked
out in I.  
$G^{(0)}_{mmmM}(1112)$ is given by Eq.(155) in I:
\be
G^{(0)}_{mmmM}(1112)
=\int d\bar{1} 3G^{(0)}_{mM}(1\bar{1})
(G^{(0)}_{2}(1\bar{1}))^{2}
(-i\omega_{1}(\bar{1})G^{(0)}_{mM}(\bar{1}2)
~~~.
\ee
This second-order contribution is given then by
\be
\chi^{(2)}_{2}(12)= \int d\bar{1}
\psi_{0}\frac{\phi_{0}(1)}{S_{2}^{(0)}(1)}
\frac{1}{2}iG^{(0)}_{mM}(1\bar{1})(G^{(0)}_{2}(1\bar{1}))^{2}
\omega_{1}(\bar{1})iG^{(0)}_{mM}(\bar{1}2)
~~~.
\ee

In order to carry out the integrals contributing
to $\chi^{(2)}$ it is convenient to write
\be
iG_{mM}^{(0)}(12)=\theta (t_{1}-t_{2})R(t_{1},t_{2})
\left(\frac{b_{12}}{2\pi}\right)^{d/2}
e^{-\left(\frac{b_{12}}{2}\right)r_{12}^{2}}
\ee
where we introduce the notation
\be
b_{ij}=\frac{1}{2(t_{i}-t_{j})}
\ee
and
\be
G_{2}^{(0)}(12)=R(t_{1},t_{0})R(t_{2},t_{0})
g_{0}
\left(\frac{a_{12}}{2\pi}\right)^{d/2}
e^{-\left(\frac{a_{12}}{2}\right)r_{12}^{2}}
\ee
where
\be
a_{ij}=\frac{1}{2(t_{i}+t_{j})}
~~~.
\ee

Let us first evaluate  $\chi_{2}^{(2)}(12)$ which is
somewhat simpler. Inserting these forms for the
correlation functions and response functions
into the integral for $\chi_{2}^{(2)}(1,2)$
we obtain:
\be
\chi_{2}^{(2)}(1,2)=
\psi_{0}\frac{\phi_{0}(1)}{S_{2}^{(0)}}\frac{1}{2}
\theta (t_{1}-t_{2})R(t_{1},t_{2})
R^{2}(t_{1},t_{0})g_{0}^{2}
K(12)
\ee
where
\be
K(12)=\int_{t_{2}}^{t_{1}}d\bar{t}_{1}
R^{2}(\bar{t}_{1},t_{0})
\omega_{1}(\bar{1})
\left(\frac{b_{1\bar{1}}}{2\pi}\right)^{d/2}
\left(\frac{a_{1\bar{1}}}{2\pi}\right)^{d}
\left(\frac{b_{\bar{1}2}}{2\pi}\right)^{d/2}
J(12)
\ee
and the spatial integral is given by
\be
J(12)=\int d^{d}\bar{r}_{1}
e^{-\left(\frac{b_{1\bar{1}}+2a_{1\bar{1}}}{2}\right)r_{1\bar{1}}^{2}}
e^{-\left(\frac{b_{\bar{1}2}}{2}\right)r_{\bar{1}2}^{2}}
~~~.
\ee

The spatial integral can be evaluated by completing
the square in the gaussian with the result:
\be
J(12)=\left(\frac{2\pi}{\alpha_{0}}\right)^{d/2}
e^{-\left(\frac{b_{\bar{1}2}(b_{1\bar{1}}+2a_{1\bar{1}})}
{2\alpha_{0}}\right)r_{12}^{2}}
\ee
where
\be
\alpha_{0} =b_{1\bar{1}}+b_{\bar{1}2}+2a_{1\bar{1}}
~~~.
\ee

Let us focus on the onsite correlation function
($r_{12}^{2}=0$)
\be
K(t_{1}t_{2})=\int_{t_{2}}^{t_{1}}d\bar{t}_{1}
R^{2}(\bar{t}_{1},t_{0})
\omega_{1}(\bar{1})
\left(\frac{b_{1\bar{1}}}{2\pi}\right)^{d/2}
\left(\frac{a_{1\bar{1}}}{2\pi}\right)^{d}
\left(\frac{b_{\bar{1}2}}{2\pi}\right)^{d/2}
\left(\frac{2\pi}{b_{1\bar{1}}+b_{\bar{1}2}+2a_{1\bar{1}}}\right)^{d/2}
\nonumber
\ee
\be
=\left(\frac{1}{2\pi}\right)^{3d/2}
\int_{t_{2}}^{t_{1}}d\bar{t}_{1}
R^{2}(\bar{t}_{1},t_{0})
\omega_{1}(\bar{1})W_{0}^{d/2}
\ee
where
\be
W_{0}=\frac{b_{1\bar{1}}a_{1\bar{1}}^{2}b_{\bar{1}2}}
{b_{1\bar{1}}+b_{\bar{1}2}+2a_{1\bar{1}}}
\nonumber
\ee
\be
=\left[8(t_{1}+\bar{t}_{1})
\left((t_{1}-t_{2})(t_{1}+\bar{t}_{1})
+2(\bar{t}_{1}-t_{2})(t_{1}\bar{t}_{1})\right)\right]^{-1}
~~~.
\ee
We have from the zeroth-order theory
\be
S_{2}^{(0)}(1)=A_{0}t_{1}
\ee
and 
\be
\omega_{1}(1)=\frac{\omega}{A_{0}t_{1}^{2}}
\ee
where $\omega$ is given at zeroth order by Eq.(\ref{eq:94}).
We have then
\be
K(t_{1}t_{2})=\frac{\omega}{A_{0}t_{1}^{2}}
\frac{1}{8^{d/2}}\left(\frac{1}{2\pi}\right)^{3d/2}
\tilde{K}(t_{1},t_{2})
\ee
where
\be
\tilde{K}(t_{1},t_{2})=\int_{t_{2}}^{t_{1}}
\frac{d\bar{t}}{\bar{t}}\left(\bar{t}\tilde{W}\right)^{d/2}
\ee
\be
\tilde{W}=8W_{0}
\ee
Using the result
\be
g_{0}=A_{0}t_{0}^{2\omega}(8\pi )^{d/2}
\label{eq:133}
\ee
which follows from Eq.(\ref{eq:95}),
and collecting coefficients, we  have
\be
\chi_{2}^{(2)}(t_{1},t_{2})=
\psi_{0}\frac{\phi_{0}(1)A_{0}}{S_{2}^{(0)}(t_{1})}
\frac{\omega}{2\pi^{d/2}}
\theta (t_{1}-t_{2})R(t_{1},t_{2})
t_{1}^{2\omega}\tilde{K}(t_{1},t_{2})
~~~.
\ee

Looking at the time integral we change variables from
$\bar{t}$ to $\bar{t}=yt_{1}$
and introduce $s =t_{2}/t_{1}$ so that
\be
\tilde{K}(t_{1},t_{2})=\int_{s}^{1}
\frac{dy}{y}\left(\frac{y}{1+y}\right)^{d/2}
[t_{1}^{2}(1-s)(1+y)+2t_{1}^{2}(1-y)(y-s)]^{-d/2}
\nonumber
\ee
\be
=t_{1}^{-d}\kappa (s)
\ee
where
\be
\kappa (s)=\int_{s}^{1}\frac{dy}{y}\left(\frac{y}{1+y}\right)^{d/2}
[(1-s)(1+y)+2(1-y)(y-s)]^{-d/2}
~~~.
\label{eq:141}
\ee
Then the second-order correction  is given by
\be
\chi_{2}^{(2)}(t_{1},t_{2})=\psi_{0}\phi_{0}(1)
\frac{\omega}{2\pi^{d/2}}
\theta (t_{1}-t_{2})R(t_{1},t_{2})
t_{1}^{2\omega -2-d}\kappa (s)
\nonumber
\ee
\be
=\psi_{0}\phi_{0}(1)\frac{\omega}{2\pi^{d/2}}
\theta (t_{1}-t_{2})R(t_{1},t_{2})
t_{1}^{-d/2}\kappa (s)
~~~.
\ee
The zeroth order contribution is given by
\be
\chi_{0}(t_{1},t_{2})=\psi_{0}\phi_{0}(1)
\theta (t_{1}-t_{2})R(t_{1},t_{2})
\left(4\pi (t_{1}-t_{2})\right)^{-d/2}
~~~.
\ee

We can then combine the zeroth and $\chi_{2}^{(2)}$
contributions to find
\be
\tilde{\chi} (t_{1},t_{2})
=\chi_{0}(t_{1},t_{2})+\chi_{2}^{(2)}(t_{1},t_{2})
=\chi_{0}(t_{1},t_{2})\left(1+\Delta_{d}(s)\right)
\ee
where
\be
\Delta_{d}(s)=\omega 2^{d-1}(1-s)^{d/2}\kappa (s)
~~~,
\label{eq:140}
\ee
and $\kappa (s)$ is given by Eq.(\ref{eq:141}).
Analytically one can show
$\Delta_{d}(0)\rightarrow\frac{\omega}{2d}$
as  $d\rightarrow\infty$,
and  $\Delta_{d}(s)\rightarrow \frac{\omega}{2}(1-s)$
as $s \rightarrow 1$.

We turn next to the evaluation of $G_{mM}^{(2)}$.
Inserting the response and correlation functions
and using the properties of the $R(t_{1},t_{2})$ we have
\be
G_{mM}^{(2)}(12)=-i\frac{1}{(2\pi )^{3d/2}}
\theta (t_{1}-t_{2}) R(t_{1},t_{2})
\int_{t_{2}}^{t_{1}}d\bar{t}_{1}
\int_{t_{2}}^{\bar{t}_{1}}d\bar{t}_{2}
\frac{g_{0}^{2}}{2}\omega_{1}(\bar{1})
\omega_{1}(\bar{2})
\nonumber
\ee
\be
\times\left(b_{1\bar{1}}
b_{\bar{1}\bar{2}}b_{\bar{2}2}\right)^{d/2}
a_{\bar{1}\bar{2}}^{d}R^{2}(\bar{t}_{1},t_{0})
R^{2}(\bar{t}_{2},t_{0}) J_{mM}
\ee
where
\be
J_{mM}=\int \frac{d^{d}\bar{r}_{1} d^{d}\bar{r}_{2}}
{(2\pi )^{d}}
e^{-\frac{b_{1}}{2}r_{1\bar{1}}^{2}}
e^{-\frac{b_{2}}{2}r_{\bar{1}\bar{2}}^{2}}
e^{-\frac{b_{3}}{2}r_{\bar{2}2}^{2}}
=\frac{e^{-\frac{\alpha}{2}r_{11}^{2}}}
{{\cal D}^{d/2}}
\ee
where
\be
b_{1}=b_{1\bar{1}}
\ee
\be
b_{2}=b_{\bar{1}\bar{2}}+2a_{\bar{1}\bar{2}}
\ee
\be
b_{3}=b_{\bar{2}2}
\ee
\be
{\cal D}=b_{1}b_{2}+b_{2}b_{3}+b_{3}b_{1}
\ee
and
\be
\alpha =\left(b_{1}^{-1}+b_{2}^{-1}+b_{3}^{-1}\right)^{-1}
~~~.
\ee
Again restricting ourselves to the onsite
response function
\be
G_{mM}^{(2)}(0,t_{1},t_{2})=-i\frac{1}{(2\pi )^{3d/2}}
\theta (t_{1}-t_{2}) R(t_{1},t_{2})
\int_{t_{2}}^{t_{1}}d\bar{t}_{1}
\int_{t_{2}}^{\bar{t}_{1}}d\bar{t}_{2}
\frac{g_{0}^{2}}{2}\omega_{1}(\bar{1})
\omega_{1}(\bar{2})
\ee
\be
\times\left(b_{1\bar{1}}
b_{\bar{1}\bar{2}}b_{\bar{2}2}\right)^{d/2}
a_{\bar{1}\bar{2}}^{d}R^{2}(\bar{t}_{1},t_{0})
R^{2}(\bar{t}_{2},t_{0}) 
\frac{1}{{\cal D}^{d/2}}
~~~.
\ee
After grouping the various multiplicative
terms, using Eq.(\ref{eq:133}), gives
\be
G_{mM}^{(2)}(0,t_{1},t_{2})
=-i\theta (t_{1}-t_{2}) R(t_{1},t_{2})
\omega^{2}\frac{2^{2d-1}}{(2\pi )^{d/2}}
\hat{J}(t_{1},t_{2})
\ee
where
\be
\hat{J}(t_{1},t_{2})=
\int_{t_{2}}^{t_{1}}d\bar{t}_{1}
\int_{t_{2}}^{\bar{t}_{1}}d\bar{t}_{2}
(\bar{t}_{1}\bar{t}_{2})^{2\omega -2}
\left(W\right)^{d/2}
\label{eq:151}
\ee
with
\be
W=\frac{b_{\bar{1}\bar{2}}a_{\bar{1}\bar{2}}^{2}}
{1+b_{2}b_{3}^{-1}+b_{2}b_{1}^{-1}}
~~~.
\ee
After some algebra we find
\be
W=\frac{1}{8(\bar{t}_{1}+\bar{t}_{2})}
\frac{1}{[(t_{1}-t_{2})(3\bar{t}_{1}-\bar{t}_{2})
-2(\bar{t}_{1}-\bar{t}_{2})^{2}]}
~~~.
\ee
After inserting this result for $W$ back into
Eq.(\ref{eq:151}), using $\omega =d/2+1$, which
is valid at this order, and letting
$\bar{t}_{1}=(t_{1}-t_{2})y_{1}$ and
$\bar{t}_{2}=(t_{1}-t_{2})y_{2}$ we obtain
\be
\hat{J}(t_{1},t_{2})=
\frac{1}{[8(t_{1}-t_{2})]^{d/2}} g(\tau)
\ee
where
\be
g(\tau)=\int_{\tau}^{1+\tau}\frac{dy_{1}}{y_{1}}
\int_{\tau}^{y_{1}}\frac{dy_{2}}{y_{2}}
\left(\frac{y_{1}y_{2}}{y_{1}+y_{2}}\right)^{d/2}
\frac{1}{[3y_{1}-y_{2}-2(y_{1}-y_{2})^{2}]^{d/2}}
\ee
and
\be
\tau=\frac{t_{2}}{t_{1}-t_{2}}=\frac{1}{t_{1}/t_{2}-1}
~~~.
\label{eq:156}
\ee
A more useful form for $g(\tau)$ is obtained
if we let $y_{2}=y_{1}z$ to obtain
\be
g(\tau)=\int_{\tau}^{1+\tau}\frac{dy_{1}}{y_{1}}
\int_{\tau /y_{1}}^{1}dz f(z,y_{1})
\label{eq:158}
\ee
where
\be
f(z,y_{1})=
\frac{z^{d/2-1}}{[(1+z)((3-z)-2y_{1}(1-z)^{2})]^{d/2}}
~~~.
\ee

This means that the MSR on site response function can be
written up to second order
\be
G_{mM}(0,t_{1},t_{2})
=-i\theta (t_{1}-t_{2})\frac{R(t_{1},t_{2})}
{[4\pi (t_{1}-t_{2})]^{d/2}}
\left(1+\omega^{2}2^{d-1}g(\tau)\right)
~~~.
\ee

Pulling together all of the results for the
physical response function function we have up
to second order
\be
\chi (t_{1},t_{2})=t_{2}^{-1-a}f(t_{1}/t_{2})
~~~,
\ee
with
\be
a=\frac{1}{2}(d-1)
\ee
and
\be
f(x)=\sqrt{\frac{2}{\pi A_{0} }}
\frac{x^{\omega -1/2}}{[4\pi (x-1)]^{d/2}}
\left(1+\Delta_{d}(s)+\omega^{2}2^{d-1}g(\tau)\right)
\label{eq:163}
\ee
where $x=t_{1}/t_{2}$.
It is not difficult to show that in the 
small $\tau$ limit 
\be
g(\tau)=-K_{d}~ln ~\tau 
+\bar{g}(\tau )
\label{eq:164}
\ee
where
\be
K_{d}=\int_{0}^{1}dz \frac{z^{d/2-1}}
{[(1+z)(3-z)]^{d/2}}
~~~.
\label{eq:165}
\ee
$K_{d}$ also appeared in the analysis in I.  
If our
perturbation theory results are to make sense
we must exponentiate the $-K_{d}~ln ~\tau$
contribution:
\be
1-v ~ln ~\tau \approx \tau^{-v}=(x-1)^{v}
\ee
where in the last teerm we have used
Eq.(\ref{eq:156}) and
\be
v=\omega^{2}2^{d-1}K_{d}
\label{eq:167}
~~~.
\ee
Putting this result back into Eq.(\ref{eq:163})
we obtain our final result for
the scaling function given by Eq.(\ref{eq:17}).

\section{Conclusions}
We have used perturbation theory to explore the nature of the 
scaling solutions for the local scalar order parameter
response function.  At lowest, gaussian level, order
we find agreement with previous work.  Going to the next
order we find that the scaling index $a$, defined by
Eq.(\ref{eq:1.5}), is unchanged from its gaussian level value
of $a=(1/2)(d-1)$.  The non equilibrium exponents
$\lambda$ and $\lambda_{R}$ are found to be equal at
lowest and second order in perturbation theory
with explicit values given as a function of $d$.
Expressions for the scaling function are also available.

It is not at all clear if there is a general proof
that $\lambda =\lambda_{R}$.  The validity of this
result at second order in perturbation theory is
very suggestive.  It will also be interesting to see
whether, within this perturbation theory, this result
holds for the case\cite{MAZ2} of vector order parameters
$(n > 1)$ where the analysis is more involved.

Acknowledgements: This work was supported
by the National Science Foundation under Contract
No. DMR-0099324.  I thank Professor Marco Zannetti
for useful discussions.


\begin{thebibliography}{99}
\bibitem{review}  A brief review is given later in this section.
\end{thebibliography}
\end{document}